\documentclass[12pt,preprint,flushrt]{aastex}
\doublespace

\begin{document}
\newcommand{\etal}{{\it et al.}}
\include{idl}

\title{Burst Oscillation Periods from 4U 1636--53: A Constraint on the 
Binary Doppler Modulation}
\author{A. B. Giles$^{1,2}$, K. M. Hill$^{1}$, T. E. Strohmayer$^{3}$,  N. 
Cummings$^3$}
\affil{
       $^{1}$ School of Mathematics and Physics\\
              University of Tasmania \\
	      GPO Box 252-21, Hobart \\
              Tasmania 7001, Australia \\
       $^{2}$ Spurion Technology Pty. Ltd.\\
              200 Mt. Rumney Road, Mt. Rumney\\
              Tasmania 7170, Australia \\
       $^{3}$ Laboratory for High Energy Astrophysics, Mail Code 662 \\ 
              NASA Goddard Space Flight Center\\
              Greenbelt, MD 20771, USA\\ }

\begin{abstract}
The burst oscillations seen during Type I X-ray 
bursts from low mass X-ray binaries (LMXB) typically evolve in period
towards an asymptotic limit that likely reflects the spin of the 
underlying neutron star. If the underlying period is stable enough, measurement
of it at different orbital phases may allow a detection of the Doppler 
modulation caused by the motion of the neutron star with respect to the 
center of mass of the binary system. Testing this hypothesis 
requires enough X-ray 
bursts and an accurate optical ephemeris to determine the binary phases at 
which they occurred. We present here a study of the distribution of asymptotic 
burst oscillation periods for a sample of 26 bursts from 4U 1636-53 observed 
with the Rossi X-ray Timing Explorer (RXTE). The burst sample includes both 
archival and proprietary data and spans more than 4.5 years. 
We also present new optical light curves of V801 Arae, the 
optical counterpart of 4U 1636-53, obtained during 1998-2001. We use these
optical data to refine the binary period measured by Augusteijn et al. (1998) 
to 3.7931206(152) hours. We show that a subset of $\sim 70\%$ of 
the bursts 
form a tightly clustered distribution of asymptotic periods consistent with
a period stability of $\sim 1 \times 10^{-4}$. The tightness of this 
distribution, made up of bursts spanning more than 4 years in time, suggests 
that the underlying period is highly stable, with a time to change 
the period of $\sim 3 \times 10^{4}$ yr. This is comparable to 
similar numbers derived for X-ray pulsars. 
We investigate the period and orbital phase data for our burst sample 
and show that it is consistent with binary motion of the neutron star with 
$v_{ns} \sin i < 55$ and $75$ km s$^{-1}$ at $90$ and $99\%$ confidence,
respectively. We use this limit as well as previous radial velocity data to
constrain the binary geometry and component masses in 4U 1636-53. Our results
suggest that unless the neutron star is significantly more massive than 
1.4 $M_{\odot}$ the secondary is unlikely to have a mass as large as 
0.36 $M_{\odot}$, the mass estimated assuming it is a main sequence star which 
fills its Roche lobe. We show that a factor of 2-3 increase in the number of 
bursts with asymptotic period measurements should allow a detection of the 
neutron star velocity. 
\end{abstract}


\section{Introduction}

Millisecond oscillations in the X-ray brightness of thermonuclear X-ray bursts
(so called ``burst oscillations'') have now been reported for 10 low mass 
X-ray binary (LMXB) systems (see Strohmayer 2001 for a review).  All of these
results are based on observations with the Proportional Counter Array (PCA) 
on the Rossi X-ray Timing Explorer (RXTE) except for the evidence for burst 
oscillations from the accreting millisecond pulsar SAX J1808-369 which is based
on SAX Wide Field Camera (WFC) data (see in't Zand et al. 2001).  
A large body of evidence supports the hypothesis that these oscillations are 
produced by rotational modulation of a hot spot (or possibly a pair of hot 
spots) induced on the neutron star surface by inhomogeneous nuclear burning. 
In particular, the large modulation amplitudes, high coherence and long term
stability of the frequency are fully consistent with the rotational 
modulation scenario (see Strohmayer, Zhang \& Swank 1997; Strohmayer et al. 
1998a; Strohmayer \& Markwardt 1999; Muno et al. 2000 and Strohmayer et al. 
1998b). 

The oscillation frequency during a burst is usually not constant.
Often the frequency is observed to increase by $\approx 1 - 3$ Hz in the 
cooling tail, reaching a plateau or asymptotic limit. 
Strohmayer et al. (1997) have suggested that the time evolution of the burst 
oscillation frequency results from angular momentum conservation of the 
thermonuclear shell. The burst expands the shell, increasing its rotational 
moment of inertia and slowing its spin rate. Near burst onset the shell is 
thickest and thus the observed frequency lowest. The shell spins back up as 
it cools and recouples to the underlying neutron star. Cumming \& Bildsten 
(2000) studied this mechanism in some detail and concluded that it appeared to 
be viable. However, more recent work by Cumming et al (2001) which corrects an 
error in the previous work and includes general relativistic effects suggests
that it may not be able to account for all of the observed frequency evolution.
Spitkovsky, Levin \& Ushomirsky (2001), however, suggest that geostrophic 
effects due to the coriolis force may account for the additional frequency 
evolution. Nevertheless, this scenario suggests that the limiting frequency 
is the neutron star spin frequency. We note, however,
that not all bursts exhibit this behavior. For example, Strohmayer (1999) and 
Miller (2000) identified a burst from 4U 1636-53 (burst 4 in Table 2) with 
a spin down of the oscillations in the decaying tail. This burst also had 
an unusually long decaying tail which may have been related to the spin 
down episode. Muno et al. (2000) also reported an episode of spin down 
in a burst from KS 1731-260. 

The long term (over year timescales) stability of burst oscillations from 
4U 1728-34 and 4U 1636-53 has been studied by Strohmayer et al. (1998b). 
For three bursts from 4U 1728-34 separated in time
by $\approx 1.6$ years they found the 363 Hz burst frequency to be highly 
stable, with an estimated time scale to change the oscillation period of about 
23,000 years. Based on a study of three bursts from 4U 1636-53 (bursts 
number 1, 2 \& 3 in Table 2) spanning a much 
shorter time interval (about 1 day) they suggested that the observed changes 
in the limiting frequency of the 581 Hz oscillation might be due to orbital 
motion of the neutron star, which could provide a way of deriving or 
constraining the X-ray mass function of the system. However, with only three 
bursts available at the time it was not possible to test this hypothesis 
definitively nor draw any strong conclusions on the mass function. 
4U 1636-53 is perhaps the best system in which to search for such an 
effect since the orbital period is known and a large sample of bursts have now
been obtained with RXTE. For plausible system parameters and the orbital 
period of $\sim$3.8 hours the expected Doppler shifts are of order a part in 
$10^{-4}$. 

The optical counterpart of 4U 1636-53, V801 Arae, has been observed 
many times since its identification in 1977 (McClintock et al. 1977) and 
a collection of photometric data, from July 1980 to May 1988, was compiled 
by van Paradijs et al. (1990) (see references therein). The van Paradijs 
et al. (1990) ephemeris was later revised by Augusteijn et al. (1998) 
who identified a cycle miss count by reanalysing all the old data 
and incorporating newer observations made between June 1992 and August 1993. 
Augusteijn et al. (1998) also reported some spectroscopic measurements of 
emission and absorbtion line features. 

In this paper we report new photometric light curves of 4U 1636-53 obtained
over the period 1998 March to 2001 May and use them to revise the ephemeris 
of Augusteijn et al. (1998). We then use this new emphemeris to derive the 
binary phases of RXTE X-ray bursts and examine the possibility that 
the distribution of observed asymptotic burst oscillation periods is 
consistent with Doppler modulation caused by the orbital velocity of the 
neutron star. The paper is organized as follows. In \S 2 we begin with
a discussion of our new optical observations. We then explore in \S 3 
the implications of our new observations for the ephemeris of maximum light 
from V801 Arae. We show that our data suggest a small correction to the orbital
period of Augusteijn et al. (1998). In \S 4 we describe the sample of X-ray 
bursts from 4U 1636-53 and we study in detail the observed distribution of 
asymptotic burst oscillation periods. We show that a subset of $\approx 70\%$
of bursts with asymptotic period measurements form a tightly clustered
distribution consistent with having been generated by a highly stable 
underlying period. We then fit this distribution to models of the period - 
phase distribution expected from binary motion of the neutron star and show 
that it is consistent with circular orbital motion of the neutron star with 
$v \sin i < 55$ km s$^{-1}$ ($90\%$ confidence). In \S 5 we summarize our 
findings and discuss their implications for the component masses and 
binary geometry of 4U 1636-53. 
We conclude with a discussion of future improvements to our constraints 
expected from a larger sample of X-ray bursts. 

\section{Optical Observations}

All the optical observations described in this paper were made using 
the Mt. Canopus 1-m telescope at the University of Tasmania observatory.
The observations used standard {\it V\/} \& {\it I\/} filters 
and the CCD reduction procedure was identical to that described in 
Giles, Hill \& Greenhill (1999). All times presented in this paper 
have been corrected to Heliocentric Julian Dates (HJD) and a complete 
journal of the observations is given in Table 1. 
For the 1998 observations the telescope was equipped with an SBIG CCD 
camera having 375 x 242 pixels with an image scale of $0.42\arcsec$ 
$\times$ $0.49\arcsec$ pixel$^{-1}$. On the nights of 1998 March 25 \& 27  
continuous pairs of {\it V\/} \& {\it I\/} integrations were obtained 
but the {\it I\/} band data are not discussed further in this paper. 
Three {\it V\/} band light curves from 1998 are shown in Figure 1 which 
plots the differential magnitudes with respect to a brighter star that 
can be located on the finder chart in McClintock et al. (1977). This 
secondary standard is at the western end of the 20$\arcsec$ scale bar 
(see Figure 2 on their 2S1636-536 chart). 4U 1636-53 is star number 3 on this 
same chart and is $\sim1.8$ {\it V\/} magnitudes dimmer than our secondary 
standard. For the 1999 and later observations the telescope was equipped 
with an SITe CCD camera having 512 x 512 pixels with an image scale of 
$0.42\arcsec$ pixel$^{-1}$. The reduction procedures for these observations 
were similar to the 1998 data and the same local secondary standard was used. 
In Figure 2 we show the light curves for the nights of 1999 June 9 and 
2001 May 7 \& 8. We do not show plots for the remaining nights listed 
in Table 1 since the individual time spans are rather limited.

\section{Optical Ephemeris}

The ephemeris for maximum optical light given by Augusteijn et al. 
(1998) is HJD = 2446667.3183(26) $\pm$ [ N $\times$ 0.15804738(42) ] where 
the errors are indicated in the round brackets. This ephemeris was based 
on observations made between 1980 July 11 and 1993 July 12 and covers a 
total of 30048 binary periods. The predictions for this ephemeris 
are shown on Figures 1 \& 2 as the dotted trace in the lower section of 
each light curve panel. There is a small phase shift evident between our 
data and the prediction after extrapolating forward in time for the 
additional $\sim$18080 binary periods to 2001 May. We have fitted a sine 
curve to the 2001 May 7 \& 8 data and determined that the phase difference 
at this epoch is $\sim0.15$. This can be eliminated by reducing the period 
of Augusteijn et. al. (1998) by a very small amount and, assuming that there 
is no cycle mis-count which is very unlikely, this period change 
corresponds to 1.65 times their quoted error. We therefore adopt the 
following new ephemeris for the time of maximum light 
HJD = 2446667.3183(26) $\pm$ [ N $\times$ 0.15804669(24) ]. 
After adjusting the modulation amplitude and mean level this ephemeris 
is plotted as the solid line in Figures 1 \& 2 and although derived from 
fitting only the data from 2001 May 7 \& 8 it appears reasonably 
consistent with the other four light curves. The night of 1998 April 3 
in Figure 1 does have an odd profile but van Paradijs et al. (1990) have 
previously commented on multi-humped profiles which they had eliminated 
from their analysis procedure. Our small change to the binary period would 
be expected to have a fairly minimal effect on the earlier light curves 
analysed by Augusteijn et. al. (1998) particularly for the older data. 
We have not attempted to revise the epoch of phase zero or its error as 
quoted by Augusteijn et al. (1998) since we do not have all the old 
raw data and phase zero is hard to define for this system where the light 
curve is quite variable and has no sharp features. In any case there is 
still an unknown relationship between the optical \& true orbit phase 
zero and this will likely remain so at least until more extensive radial 
velocity observations are available. Throughout this paper phase zero 
is defined as the optical maximum when superior conjunction of the 
companion star is thought to occur (neutron star closest to the Earth).

\section{Asymptotic Oscillation Periods of RXTE X-ray Bursts}

A total of 30 X-ray bursts from 4U 1636-53 are available to us as public 
or PI data from the PCA experiment on RXTE 
and information about them relevant to this study are listed in Table 2. 
A comprehensive description of the properties of these bursts will be given 
elsewhere (Cummings \& Strohmayer 2001). Here we will be primarily interested
in the asymptotic burst oscillation periods and inferred binary orbital phases 
of the bursts. 
The 1.72 ms (581 Hz) oscillation in most of these bursts exhibits a 
characteristic evolution towards a limiting (shortest) period in the tail of 
the burst. It was our aim to try and measure this limiting period for each 
burst in the sample. For most of these bursts we had event mode data with a 
time resolution of 1/8192 seconds across the entire 2 - 90 keV PCA bandpass. 
In a few cases we had binned data with the same time resolution. We began by 
correcting the event arrival times to the solar system barycenter using the 
JPL DE200 ephemeris and the standard RXTE analysis tools (either fxbary, or
faxbary for the most recent data). We then calculated dynamic variability 
spectra using the $Z_1^2$ statistic (see Strohmayer \& Markwardt 1999 for a 
discussion and example). Such spectra are essentially similar to standard FFT 
dynamic power spectra except that we oversample in frequency. We used 2 s 
intervals and start a new interval every 0.125 s. We oversample in frequency
by a factor of 16. For each burst we
calculated two dynamic spectra, one using data across the entire bandpass, and
a second using only a hard band from $7 - 20$ keV. We did this because burst
oscillation amplitudes are often stronger at higher energies (see for example
Strohmayer et al. 1997).  To determine the asymptotic period we searched the 
pair of dynamic power spectra of each burst and determined
the shortest period detectable during each burst. By detectable we mean that
the signal peak had to be larger than $Z_1^2 > 16$, which corresponds to a
single trial significance of $3.4 \times 10^{-4}$. As an example Figure 3 shows
a typical dynamic spectrum from one of our bursts and the power spectrum from 
which the asymptotic period was deduced (burst 20 in Table 2, in this case 
the spectrum from the hard band). In most cases a clear
frequency track of the oscillation could be seen in the dynamic power 
spectrum, and the procedure was straightforward. In several cases, either
the oscillations were very weak or the frequency evolution was ``anomalous''
(meaning the frequency was observed to decrease with time), and in these
cases we judged that an asymptotic period could not be reliably measured.
An example of this is the burst which occurred on 1996 December 
31 (burst 4 in Table 2) and has been discussed in detail by Strohmayer (1999).
We note that this was the case for only 4 bursts in our sample, so that in the
majority of cases the asymptotic period was reasonably well defined. Although 
these bursts could not be used for the present investigation, for completeness,
we also include them in Table 2. We selected the shortest asymptotic period 
measured in either power spectra as the asymptotic value for that burst. 
These periods are also listed in Table 2.

The column in Table 2 showing the burst binary phases has been derived 
using the new optical ephemeris described in the previous section. 
The phase error for each burst is dominated by the ability to determine 
the optical phase zero for any particular epoch but is typically $<\pm0.05$. 
Relative phase errors are much smaller given the $>$48,000 cycle time 
span of the optical observations and the fact that the X-ray bursts used here 
all occur within a time interval of $\sim4.4$ years (only 10,000 cycles) 
ending in 2001 May. 

\subsection{Period Measurement Uncertainty}

An important quantity to understand is the characteristic error, $\sigma_P$, 
in
our period measurements. To estimate this we have carried out a series of 
simulations which mimic the conditions of our asymptotic period measurements.
To do this we first generate a count rate model comprised of a constant plus 
a sinusoidal modulation of fixed period and amplitude. We then generate random 
realizations of this model using the same temporal resolution as our burst 
data. We model a 2 s interval of data since this was the interval length 
we used for all our dynamic spectra. We use a count rate and modulation 
amplitude typical of the intervals in the tails of bursts where we actually 
measure the asymptotic periods. We then compute the $Z_1^2$ spectra for each of
the simulated data sets and determine the centroid period of the signal. 
Since typically we follow the signal in a real burst down to or near a limiting
threshold (in this case $Z_1^2 = 16$), we only keep simulated period 
measurements for which the peak signal power was close to our limiting 
threshold. In practice we found that $16 < Z_1^2 < 24$ was characteristic of 
our actual asymptotic period measurements. We 
then determine how these simulated periods are distributed around the true
period. Specifically we fit a gaussian to the distribution of simulated 
periods and identify the width of this gaussian with the characteristic 
uncertainty, $\sigma_P$, in any one of our period measurements. Figure 4 shows 
the period distribution and best fitting gaussian derived from one of these 
simulations. We find that the typical measurement error associated with one of
our periods is $\sim 2.2\times 10^{-4}$ ms. Note that this is purely 
a statistical uncertainty. Another source of possible systematic error is 
associated with the assumption that the last period detected in a dynamic 
spectra represents a limiting value. We will have more to say on this in a 
later section.

\subsection{The Observed Distribution of Asymptotic Periods}

We used the period measurements from Table 2 to construct a distribution
of asymptotic periods. Figure 5 shows a histogram representation of the 
distribution. Although the range of all observed periods is rather large, a 
subset of $\sim 70\%$ of the bursts form a tightly peaked 
distribution. Also shown in Figure 5 is the gaussian model which best 
fits this cluster of periods.  The gaussian is centered at $1.71929 \pm
1.0 \times 10^{-4}$ ms, has a width of $2.3\times 10^{-4} \pm 
1.2 \times 10^{-4}$ and gives an excellent fit to the data. This
subset is comprised of bursts from all epochs of our sample, and suggests 
that a highly stable underlying period is responsible for this component of the
asymptotic period distribution. Note also that the width of this distribution
is comparable to our estimate above of the typical width which would be 
produced by statistical uncertainties alone. This suggests that any systematic
error associated with our measurements not reflecting a true limiting value
are small, at least within this subset of the entire sample. 

\subsection{A Constraint on the Orbital Doppler Modulation}

Assuming that the burst oscillations do reflect the spin of the neutron star
the binary motion should imprint doppler modulations on the measured periods.
We use the values from Table 2 to construct in Figure 6 a plot of 
asymptotic period against photometric (orbital) phase. Visual inspection of 
this plot reveals no strong indication of a sinusoidal modulation that 
might be produced by a sufficiently strong Doppler modulation. Such a 
modulation
would likely have a peak on Figure 6 at a phase of $\sim0.25$, when the neutron
star has a maximum recession velocity assuming that photometric maximum occurs
at superior conjunction of the secondary. We 
tested this conclusion quantitatively by fitting a period - phase model to 
the data. We used the model
\begin{equation}
P_i = P_0 \left ( 1 + (v\sin i / c) \sin (2\pi(\phi_i - 1.0)) \right ) ,
\end{equation}
where $P_0$, $v \sin i$, and $\phi_i$ are the period measured at 
inferior conjunction of the neutron star (neutron star nearest to observer),
the projected orbital velocity of the neutron star with respect to the center
of mass of the binary, and the orbital phase at
which the burst occurred, respectively. Figure 6 also shows the 
results of such fits.
The model prefers a small $v\sin i /c = 6\times 10^{-5}$, with $\chi^2 = 19.6$ 
for 16 degrees of freedom. This model is the solid curve in Figure 6. However, 
the difference between this fit and one with $v\sin i /c = 0$ is not 
statistically significant ($\Delta\chi^2 = 0.53$), hence the data are 
consistent with no doppler modulation. The $90 \%$ and 
$99\%$ confidence upper limits ($\Delta\chi^2= 2.71$ and $6.63$) on $v\sin i$ 
are 55 and 75 km $s^{-1}$, respectively. The model with $v_{ns}\sin i = 55$
km $s^{-1}$ is the dashed curve in Figure 6. Note that these fits assumed that 
the relative phase of the modulation is known based on the photometric 
ephemeris. If we relax this assumption and allow the phase of the peak 
modulation to be a parameter we find a better fit with $v\sin i = 59.3$ 
km $s^{-1}$, with $90\%$ confidence range of $15.8 < v\sin i < 102.7$ km 
$s^{-1}$ (dotted curve in Figure 6). However, the phase offset
required would be 0.2 away from that implied under the assumption that 
phase zero (photometric maximum) is at superior conjunction of the 
secondary. Although this seems large it might be possible if X-ray heating of 
the disk bulge and accretion stream interaction region contribute to the 
observed optical modulations. We discuss this further below.

Although we do not detect any doppler modulation we were able to place an upper
limit on $v\sin i$ from the period - phase data. Since there was no strong
evidence for a modulation with orbital phase we also investigated the 
upper limit using only the expected distribution of periods for 
a given $v_{ns}\sin i$ and $\sigma_P$. To do this we generated an expected 
period distribution by sampling a large number of random periods from the 
model. Samples were drawn uniformly in orbital phase and the random period was
selected from a gaussian distribution with width $\sigma_P$ centered on the 
model period for that phase. We then binned the sample periods in the 
same manner as the data and computed a $\chi^2$ goodness of fit statistic 
$\chi^2 = \sum_j (O_j - M_j)^2 / M_j $. Since our data have small numbers of 
events in each bin we computed the upper limit for $v_{ns} \sin i$ using 
monte carlo simulations.  Our resulting upper limit using this method is in 
good agreement with our result from the period versus phase fits. 

\section{Summary and Discussion}

We have investigated the asymptotic period distribution of burst oscillations
in a large sample of bursts from 4U 1636-53. We find that $\sim 70\%$ of these
bursts form a tight distribution consistent with being produced by a highly
stable mechanism such as rotation of the neutron star.  The fact that the
distribution is made up of bursts spanning a time scale of 4.4 years and 
has a characteristic width of $\Delta P / P = 1.3 \times 10^{-4}$ 
indicates that the time scale to change the underlying period is $\tau 
> \Delta T P / \Delta P = 3.4 \times 10^4$ yr. This is comparable to the 
overall period stability estimated for the 363 Hz oscillations in 4U 1728-34
(see Strohmayer et al. 1998b), and is a number characteristic of other 
rotating neutron stars such as X-ray pulsars. This provides further evidence
that rotation of the neutron star sets the burst oscillation period. 

Why do some of the bursts fall well outside this distribution?  It seems likely
that several effects may be at work here. One problem is that the
oscillation in some bursts does not remain strong enough to detect for a
long enough time interval within the burst, so that the asymptotic limit
is not reached. This results because burst oscillation properties are not
identical from burst to burst. Another possible effect was discussed by 
Cumming \& Bildsten (2000). They argued that as long as the burning shell was 
not recoupled to the neutron star the frequency observed in the burst tail 
would deviate slightly (by about 1 part in $10^{-4}$) from the neutron star 
spin frequency. This comes about because the thickness of the cooling 
atmosphere in the tail is different to the initial thickness by about 1 m, 
though the exact amount depends on the mean molecular weight of the burned 
material which in turn depends on how complete the burning was and would be 
expected to vary from burst to burst. Although this could conceivably be a 
source of additional scatter in the asymptotic periods the fact that our
observed distribution has a width comparable to that expected based on 
statistical uncertainty alone suggests that if operating at all it must be
small. If the asymmetry on the star is created by a nonradial oscillation
mode (see for example Bildsten \& Cutler 1995; Strohmayer \& Lee 1996; Heyl 
2001), then the observed oscillation frequency would always be close to the 
spin frequency or perhaps a multiple $m\Omega$ of it, but it could change by 
$\sim 1 Hz$ due to long term changes in the surface layers of the neutron 
star.
This could produce outliers in the period distribution, but would also tend to 
produce a tight component as long as surface conditions were similar for 
enough bursts. Recently, Spitkovsky, Levin \& Ushomirsky (2001) have also
studied mechanisms which can cause frequency drift. They suggest that Coriolis
forces can have an important effect and might introduce shifts in the observed
frequency comparable to those expected from radial uplift.

\subsection{Constraints on the Binary Geometry} 

In general the optical flux from LMXBs is thought to be dominated by the 
accretion disk (see van Paradijs \& McClintock 1995 for a review). There are 
three regions of a LMXB system which might contribute 
to its optical variability due to X-ray heating. These are the accretion 
disk itself, a bright spot or bulge on the outer edge of the accretion 
disk formed by interaction of the accretion stream with the disk, and the 
hemisphere of the companion facing the neutron star which is not shadowed by 
the accretion disk. In LMXBs with relatively low inclinations 
($i \lesssim 60^{\circ}$) it is this last region which is thought to dominate 
the optical modulations from the rest of the system (van Paradijs 1983, 
van Paradijs \& McClintock 1995). These systems generally produce sinusoidal
optical modulations. The optical maximum therefore occurs when 
the companion is on the far side of the neutron star (superior conjunction) 
but there may be some asymmetry or variation about the 
mean profile due to gas flows causing various X-ray shielding effects 
(Pedersen et al. 1982a).

In order to explore the implications for the binary geometry of our radial 
velocity limit for the neutron star we have created in Figure 7 a plot of the 
Roche geometry for 4U 1636-53.
For the neutron star we assumed a mass of $1.6 M_{\odot}$. For the secondary
we use a mass of $0.36 M_{\odot}$ (see Smale \& Mukai 1988; Patterson 1984).
With these masses and the known 3.8 hr orbital period the binary separation
is $\sim 1.58 R_{\odot}$. The velocities of the neutron star and secondary with
respect to the center of mass would be $91$ and $390$ km s$^{-1}$, 
respectively. Figure 7 shows a view looking down on the orbital plane of the
system. The numbers circling the system denote orbital phase positions assuming
phase zero occurs at superior conjunction of the secondary (photometric 
maximum). The dashed circle shows the extent of a disk which fills 90 \% of the
Roche lobe, a radius at which tidal effects will likely truncate it
(see for example Frank, King \& Lasota 1987). Constraints from analysis of 
optical reprocessing of X-ray bursts also indicate a large accretion
disk in 4U 1636-53 (Pedersen et al. 1982a). We also show on the plot inferred 
locations of the radial velocity components measured by Augusteijn et al. 
(1998) and given in their Table 6. Since the inferred velocity amplitudes
from their three sets of fits were all rather similar we just used the average
velocity as well as the average uncertainty. We plotted with triangles the
$\pm 1\sigma$ average velocity amplitude at the phases of superior 
conjunction given for each of their three fits. The phase locations were 
deduced by assuming that the entire binary system rotates rigidly about the 
center of mass. We also shaded the region enclosed by the triangles to further 
highlight its location. Augusteijn et al. (1998) suggested the radial velocity 
components could be identified with the bulge region associated with the 
interaction of the accretion stream with the disk. Our plot certainly 
supports this suggestion, since the shaded region is consistent with where 
the accretion stream would likely impact the disk. The location of 
the shaded region also suggests that the bulge might be a significant 
component
with regard to optical modulations. In particular, if photometric maximum 
occurs
closer to a phase of 0.8 in Figure 7, when the X-ray illuminated portion
of the bulge is facing the observer, then the implied phase shift is in the 
same
sense as that suggested by the fits to the period - phase data of the X-ray 
bursts with the phase shift left as a free parameter. More detailed 
modelling would be required to determine if the bulge can indeed effect the
optical modulations at this level, but the period - phase fits are suggestive. 

We also note that although the three simultaneous X-ray \& optical 
bursts discussed by Pedersen et al. (1982a) (see page 336) have relatively 
large error bars on the optical time delays we have re-examined them in 
the light of our new ephemeris and the system model shown in Figure 7. 
The optical delays in these bursts appear more consistent, both in delay and 
phase, with the reprocessed X-ray burst optical flux coming from the outer 
parts
of our shaded region in Figure 7 than from the facing hemisphere of 
the companion star. Although there is no evidence of a second optical 
pulse from the companion in the many optical bursts studied by Pedersen 
et al. (1982b) a weaker following pulse might easily be lost. Such a pulse
might only be evident at optimum binary phases, around phase 0.85, with 
reprocessing delays always tending to broaden and confuse the light 
curve features.

Although the radial velocities of the neutron star and secondary are not
well measured in 4U 1636-53, as Figure 7 suggests the system is rather well 
constrained. The lack of eclipses implies that $i \lesssim 76^{\circ}$. 
In addition, no dipping or partical eclipses have been observed from 
4U 1636-53. The modelling of Frank, King \& Lasota (1987) suggests that 
$i \lesssim 60^{\circ}$ in such cases. We can combine our limits on the 
velocity of the neutron star with the radial velocity measurements to place
constraints on the component masses. With the known orbital period we have 
that the neutron star velocity, 
\begin{equation}
v_{ns} < \frac{394.5 \;\; M_1 \;\; \sin i}{(M_1 + M_2)^{2/3}} \;\;\; 
{\rm km \;\;s}^{-1} ,
\end{equation} 
with $v_{ns}$ set to either our 90 or $99\%$ limit (see \S 4.3 above).
To derive mass constraints from the radial velocity data we required that
the inferred location of the radial velocity components (determined from the
velocity amplitude and phase of superior conjunction data of Augusteijn et al.
1998, see discussion above) must fit within $90\%$ of the Roche lobe radius 
of the neutron star (a likely size for the accretion disk). This further 
assumes that the entire binary rotates rigidly around the center of mass. Our 
constraints are summarized in Figure 8. We show allowed regions in 
the component mass plane for a pair of different inclinations (50 and 
$60^{\circ}$) for our $90$ and $99\%$ 
neutron star velocity limits. Indeed for $v_{ns} \lesssim 55$ km s$^{-1}$ the 
mass of the secondary must be significantly less than the $0.36 M_{\odot}$
estimate based on the mass - radius relation for main sequence stars. Further,
if the secondary is $\sim 0.3 M_{\odot}$ then the neutron star must be
quite massive $M_{ns} > 1.8 M_{\odot}$. 
The radial velocity constraints essentially exclude $i \lesssim 40^{\circ}$ 
for any reasonable masses of the components. This is because the disk cannot 
be big enough to allow high radial velocities if the inclination is too low.
Although this conclusion is dependent on our assumptions for deriving the 
radial velocity constraints, observations of large amplitude 
oscillations on the rising edge of bursts from 
this source also indicate that the inclination cannot be too low (see Nath, 
Strohmayer \& Swank 2001). These arguments suggest a likely range for the 
inclination of $50^{\circ} < i < 60^{\circ}$. With this inclination a likely 
range of masses for the neutron star and secondary are, in solar units, 
$1.4 < M_{ns} < 1.6$ and $0.2 < M_{sec} < 0.25$. More precise limits on the 
radial velocity of either component will allow more precise mass limits to be
inferred. 

Clearly additional optical photometry and spectroscopy are required for 
4U 1636-53, and at some time in the future it would prove worthwhile to 
collect together all the optical observations of 4U 1636-53 to derive a 
fully consistent ephemeris. As more burst data become available it should
become possible to measure the neutron star velocity. For example, with 
a factor of 2-3 increase in the number of bursts with reliable 
asymptotic periods and with a burst oscillation period measurement uncertainty 
of $2.2\times 10^{-4}$ ms, our simulations suggest that a velocity of 
$55$ km s$^{-1}$ (equal to our current $90\%$ upper limit) can be detected at 
better than $3\sigma$ confidence. Figure 9 shows the results of such a 
simulation for 36 burst asymptotic period measurements. 
The bursts listed in Table 2 were found in observations totaling $\sim 1.2$ 
Msec of exposure. Based on this X-ray burst rate the presently approved RXTE 
observing time on 4U 1636-53 (1.15 Msec in AO6) can be expected to provide 
another $\sim 28$ X-ray bursts, which should roughly double the sample.
Since RXTE provides much higher quality X-ray burst 
profiles than did Hakucho, further attempts to get simultaneous X-ray - 
optical
burst observations are clearly worthwhile but this requires and
is dependant on the availability of a large optical telescope.

\acknowledgments We thank Holger Pedersen for re-examining and confirming 
the dates and times of observations of 4U 1636-53 made in 1980. Archive data 
was obtained from the High Energy Astrophysics Science Archive Research 
Center Online Service provided by the NASA / Goddard Space Flight Center. 
We also thank the referees for their informative comments.

\pagebreak

\pagebreak
\centerline{\bf Figure Captions}

\figcaption[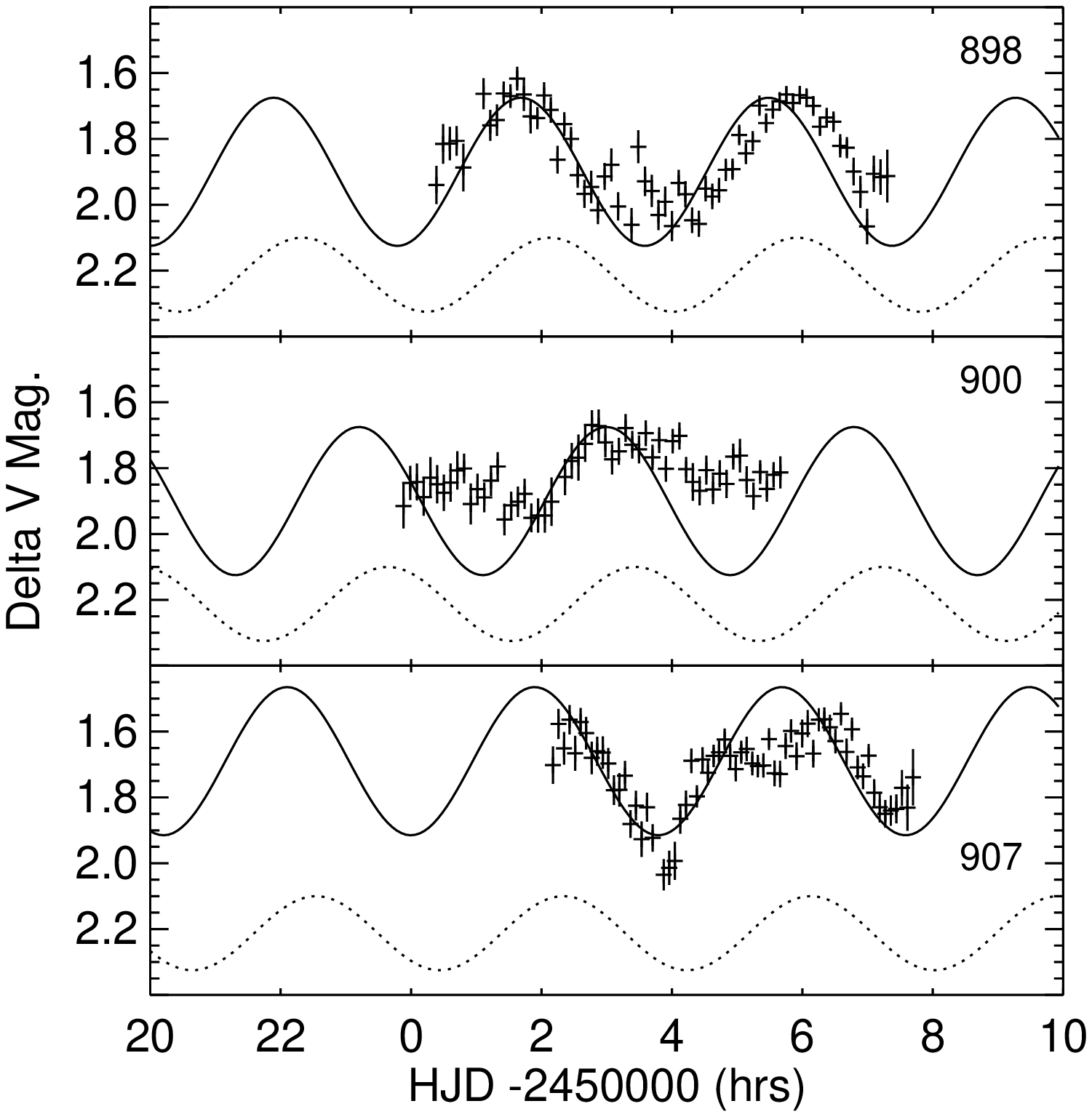]{The {\it V\/} band light curves for 4U 1636-53 on 
1998
March 25 \& 27 and 1998 April 3. The solid sine curve marks our new ephemeris 
with a period of 3.70312064 hours. The dotted curve shows the ephemeris 
prediction of Augusteijn et al. (1998) for the same three nights with an 
arbitrary offset and amplitude. The number in each panel refers to the HJD 
starting at zero hours within each light curve. \label{fig1}}

\figcaption[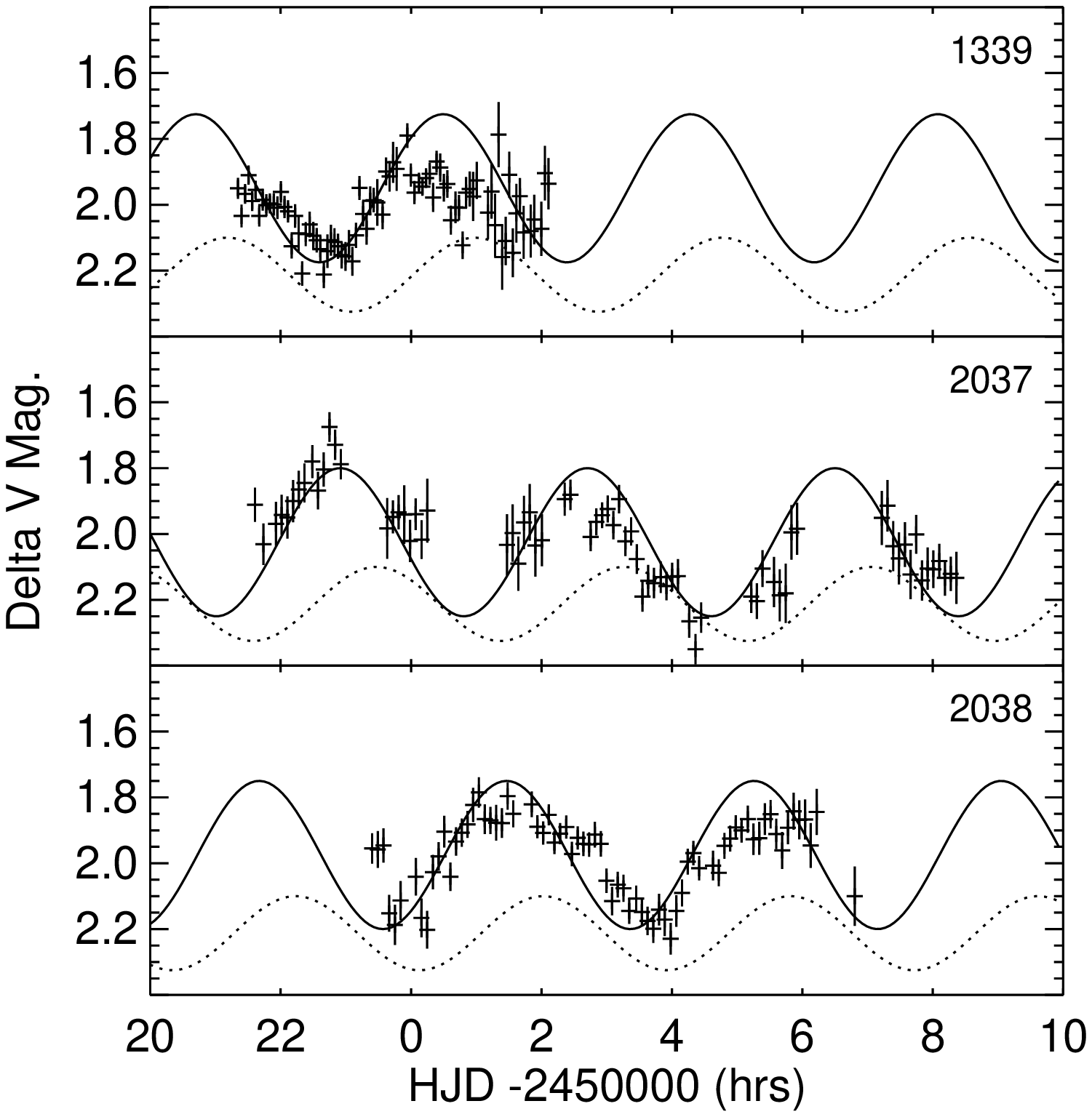]{The {\it V\/} band light curves for 4U 1636-53 on 
1999
June 9 and 2001 May 7 \& 8. The solid sine curve marks our new ephemeris 
and the first optical maximum on 7 May occurs at HJD = 2452036.954706 
The dotted curve shows the ephemeris prediction of Augusteijn et al. (1998) 
for the same three nights with an arbitrary offset and amplitude. The number 
in each panel refers to the HJD starting at zero hours within each light 
curve.\label{fig2}}

\figcaption[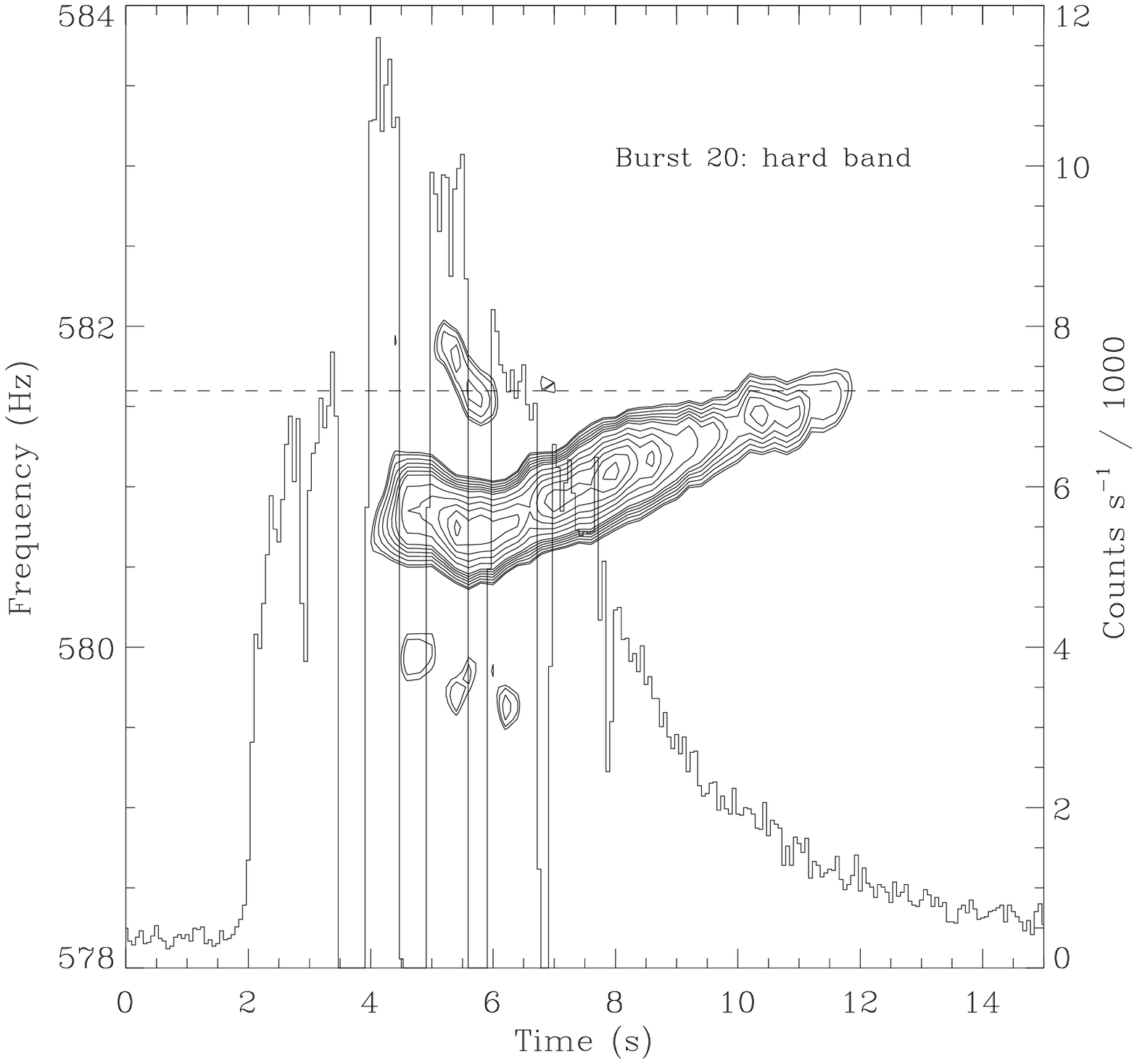]{Dynamic $Z_1^2$ spectrum in the hard X-ray
band (7 - 20 keV) for burst number 20 in Table 2 (top). The horizontal dashed 
line marks the asymptotic period inferred for this burst. The burst
lightcurve is overlaid (right axis). The gaps in the lightcurve are due to 
telemetry limitations for this data mode. Also shown is the $Z_1^2$ spectrum 
in the tail of the burst from which the asymptotic period was measured 
(bottom). In this case the vertical dashed line marks the asymptotic period. 
\label{fig3}}

\figcaption[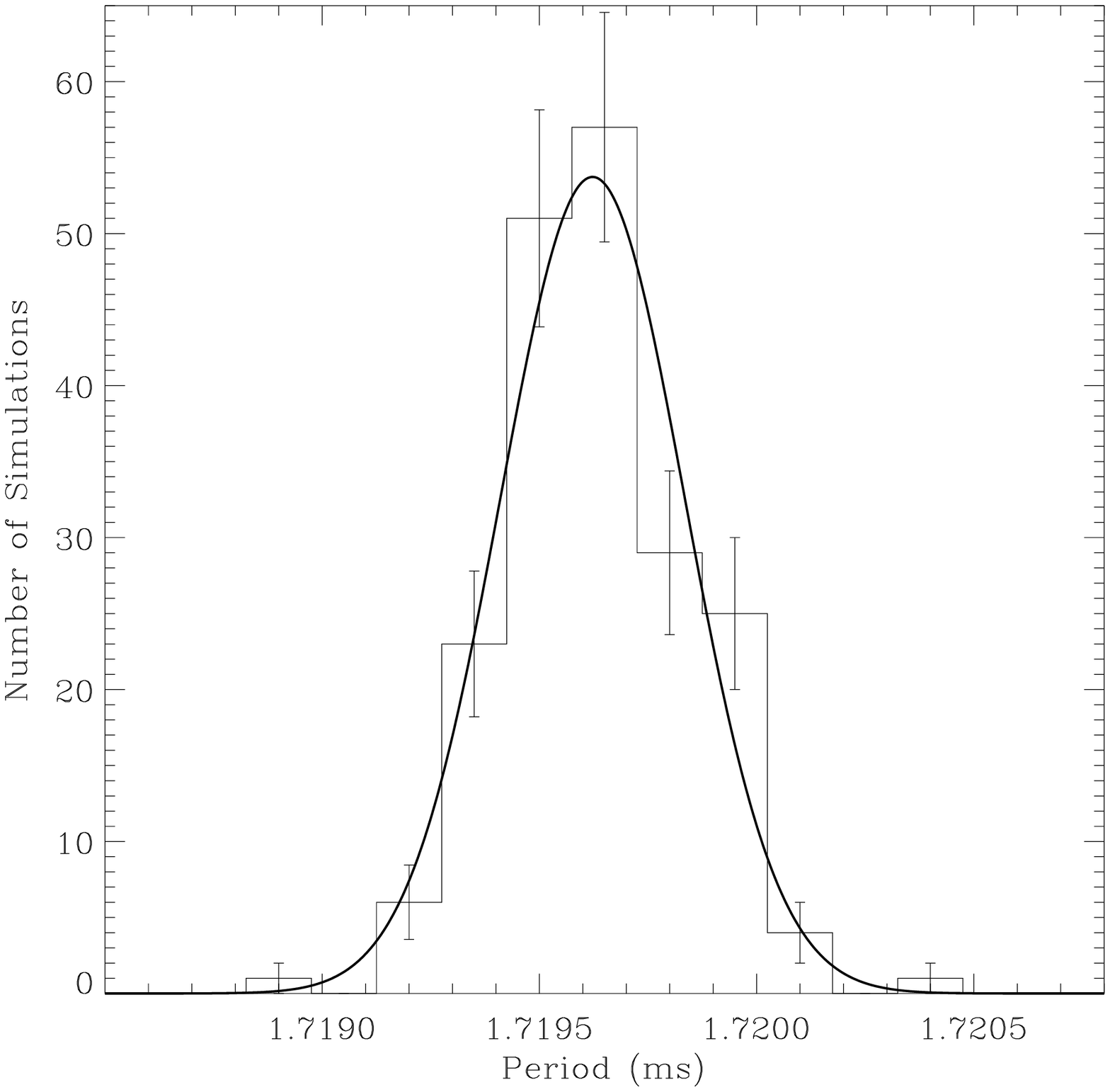]{Histogram of simulated period measurements and
best fitting gaussian distribution. See the text (\S 4.1) for a discussion of 
the simulations. The width of the gaussian is $2.2\times 10^{-4}$ ms  
and represents the characteristic uncertainty in our asymptotic period 
measurements. \label{fig4}}

\figcaption[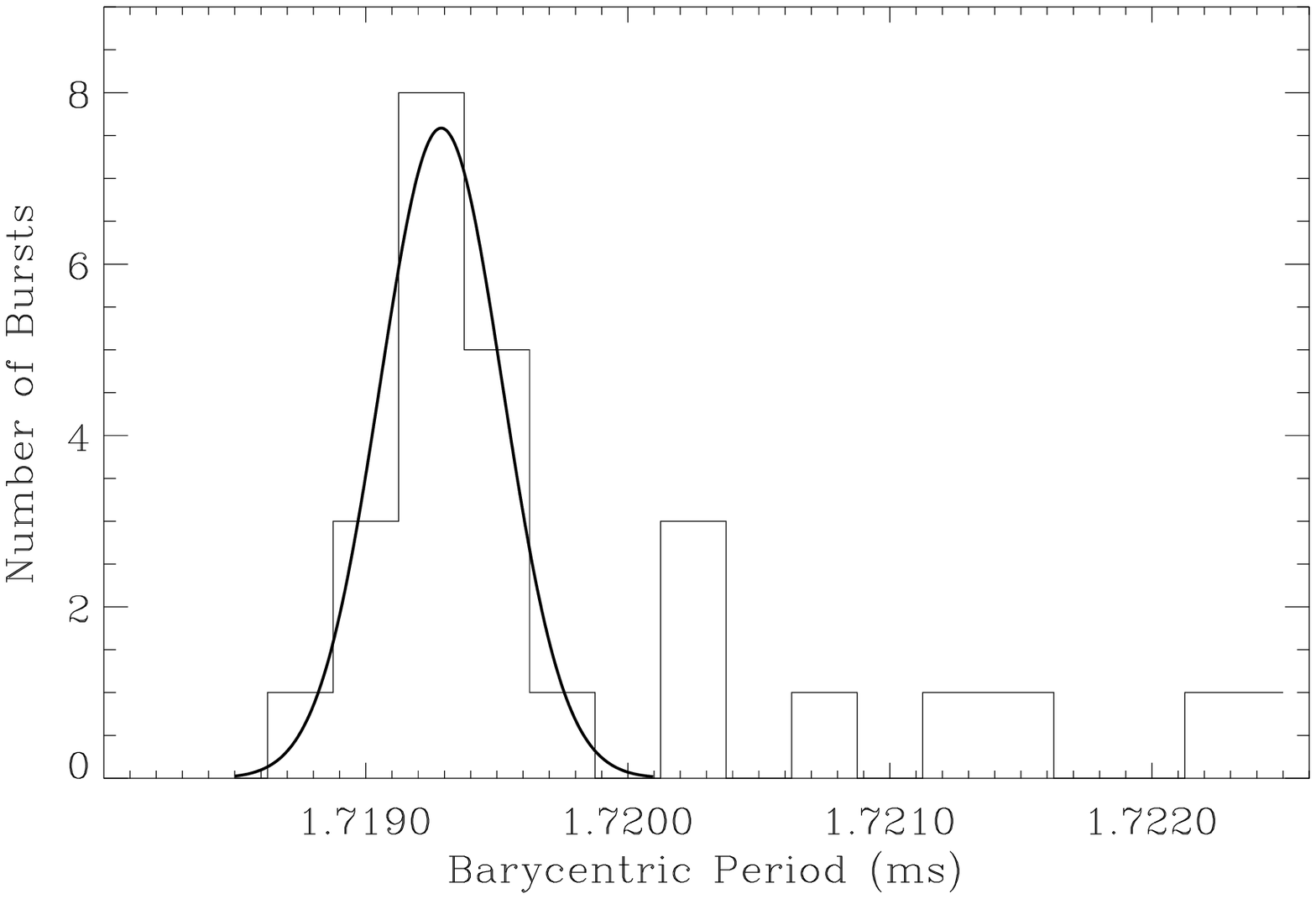]{Histogram of measured asymptotic burst 
oscillation periods for 4U 1636-53. The periods are corrected to the solar 
system barycenter. Note the cluster of 18 periods centered near 1.7192 ms. 
A gaussian distribution centered at 1.71929 ms, of width 
$\sigma = 2.3\times 10^{-4}$ ms 
fits these data well and is shown by the thick solid curve. Note the presence 
of outliers towards longer period, but none shortward of the gaussian. 
\label{fig5}}

\figcaption[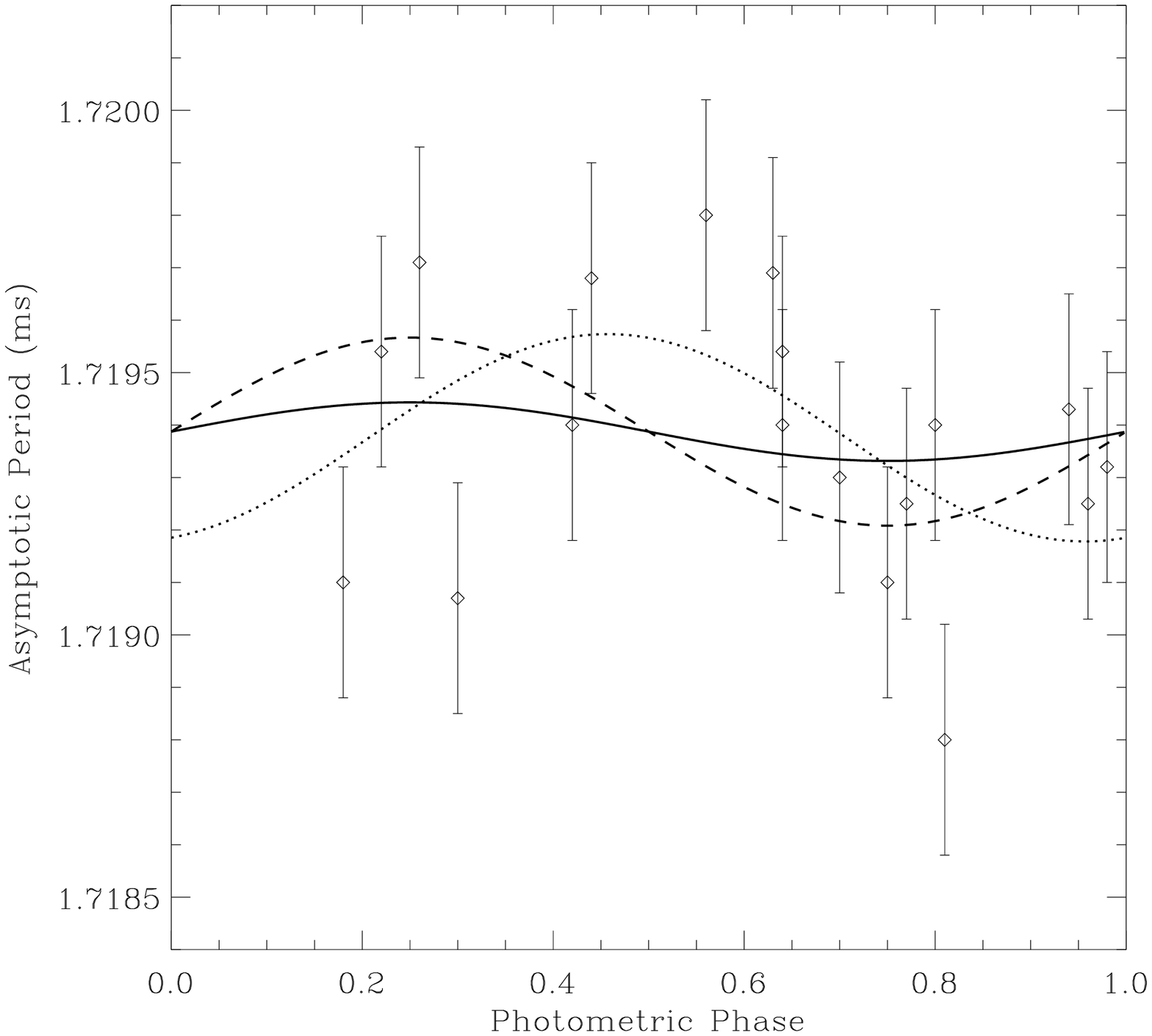]{Plot of asymptotic period versus orbital
phase for the subset of 18 bursts which have a tightly clustered period
distribution.  The solid curve is the best fitting doppler model with
$v_{ns} \sin i = 18$ km s$^{-1}$. This fit, however, is not statistically
significant compared to a $v_{ns} \sin i = 0$ model (see \S 4.3). The dashed
curve shows the model with $v_{ns} \sin i = 55$ km s$^{-1}$, which is equal to
our $90\%$ confidence upper limit. The dotted curve shows the best fitting 
model when the phase of maximum is added as an additional parameter. This fit
has $v_{ns} \sin i = 59.3$ km s$^{-1}$. \label{fig6}} 

\figcaption[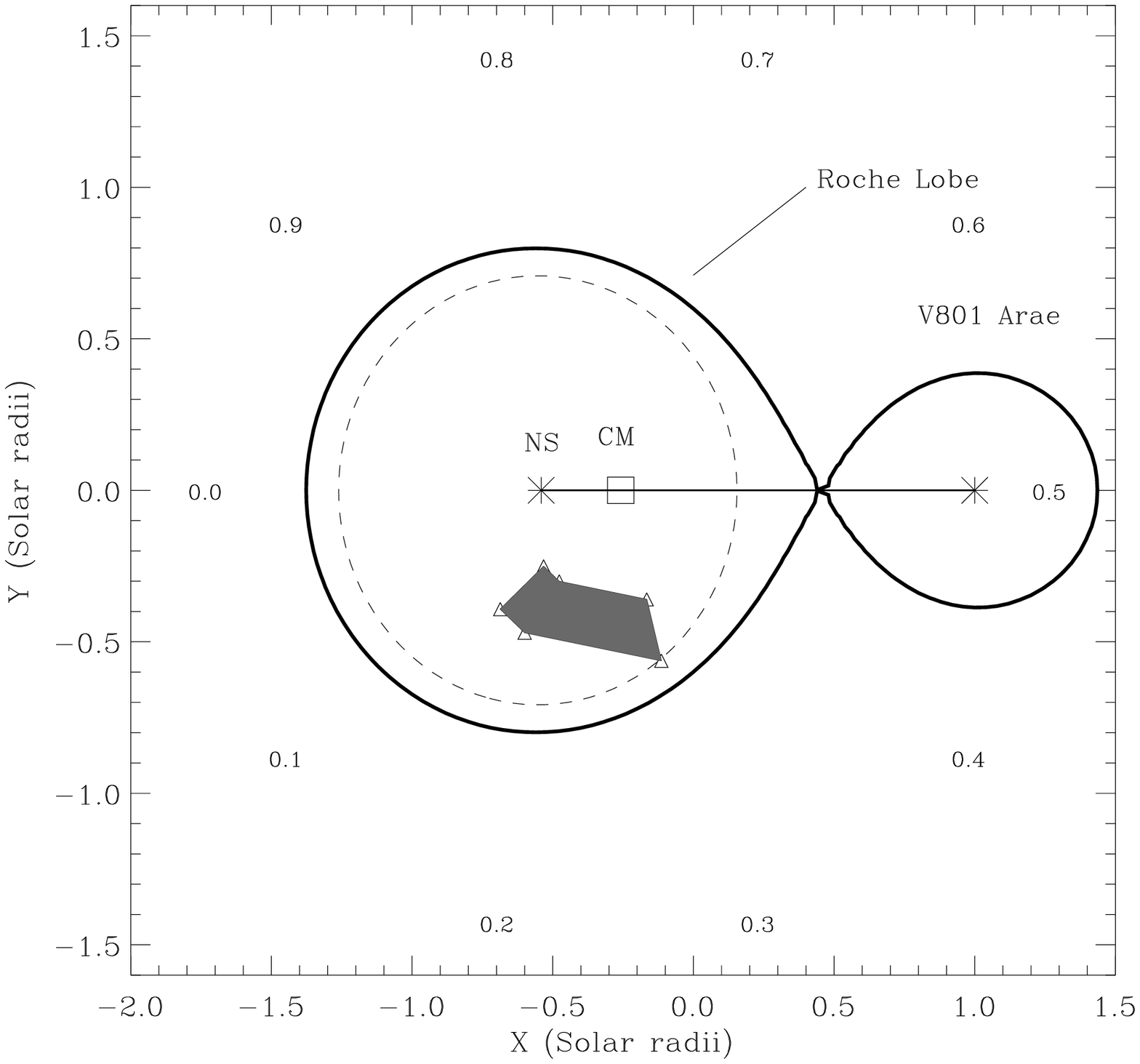]{Diagram of the Roche 
geometry for 4U 1636-53. The figure was drawn assuming neutron star and 
secondary masses of $1.6 M_{\odot}$ and $0.36 M_{\odot}$, respectively. 
The numbers circling the components correspond to orbital phases under the
assumption that phase zero corresponds to superior conjunction of the 
secondary (V801 Arae). The center of mass (CM) is denoted by a square symbol. 
The dashed circle around the neutron star marks the likely extent of an 
accretion disk under the assumption that it fills $90\%$ of the neutron star
Roche lobe. The triangles and shaded region mark the inferred locations of
the radial velocity components measured by Augusteijn et al. (1998). 
\label{fig7}}

\figcaption[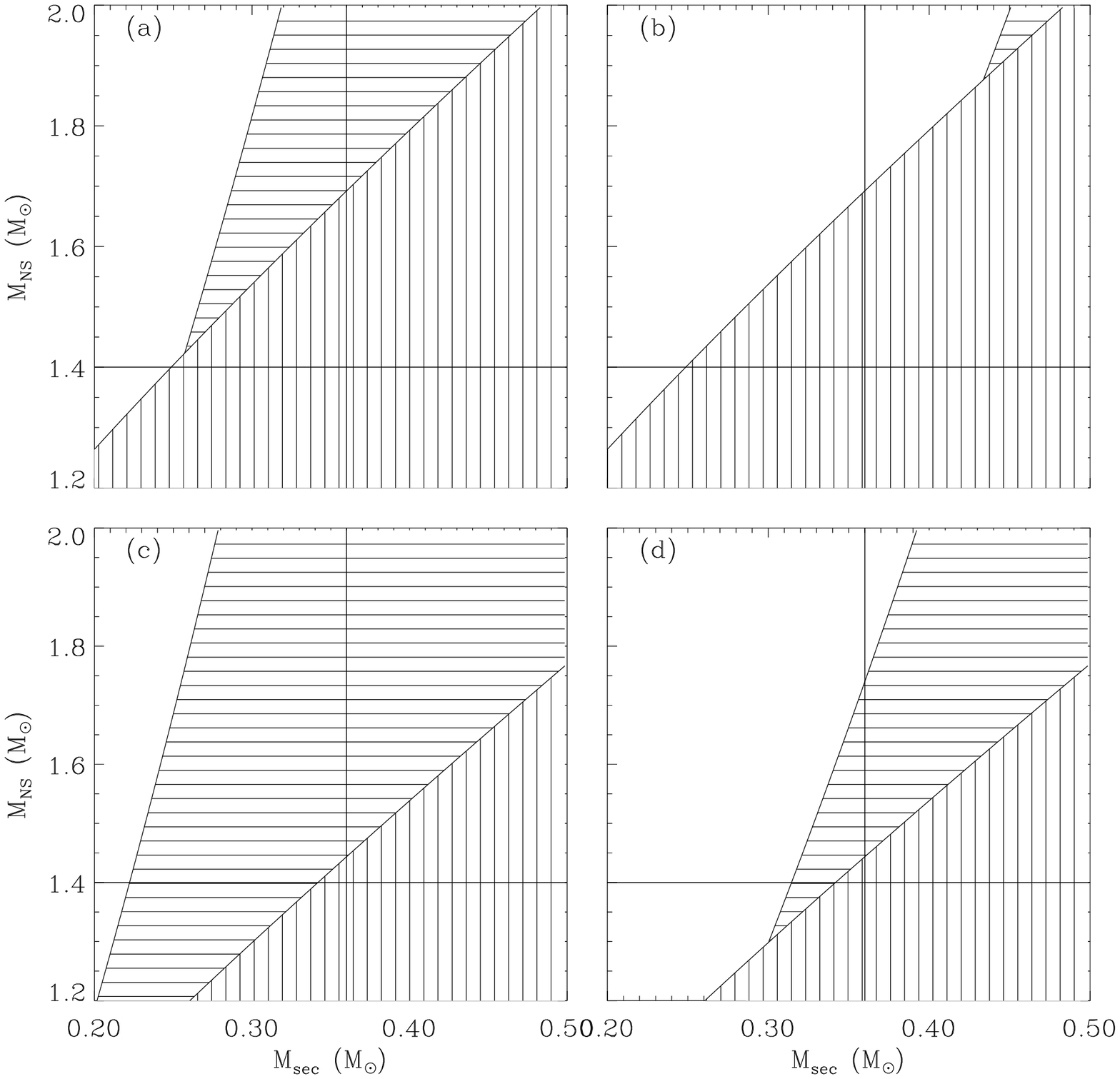]{Constraints on the component masses in 
4U 1636-53 derived from our upper limit on $v_{ns} \sin i$ and the radial
velocity data of Augusteijn et al. (1998). The regions with horizontal hatching
are excluded by the neutron star velocity limit, while the vertical hatched
regions are excluded by the radial velocity data. We show constraints for
$i = 50^{\circ}$ and $v_{ns} \sin i < 55$ km s$^{-1}$ (a), $i = 50^{\circ}$ 
and $v_{ns} \sin i < 75$ km s$^{-1}$ (b), $i = 60^{\circ}$ and $v_{ns} \sin i 
< 55$ km s$^{-1}$ (c), $i = 60^{\circ}$ and $v_{ns} \sin i < 75$ km s$^{-1}$ 
(d). The thick lines denote $M_{ns} = 1.4 M_{\odot}$ and 
$M_{sec} = 0.36 M_{\odot}$, respectively. See \S 5.1 for a discussion on how
the constraints were derived. \label{fig8}}

\figcaption[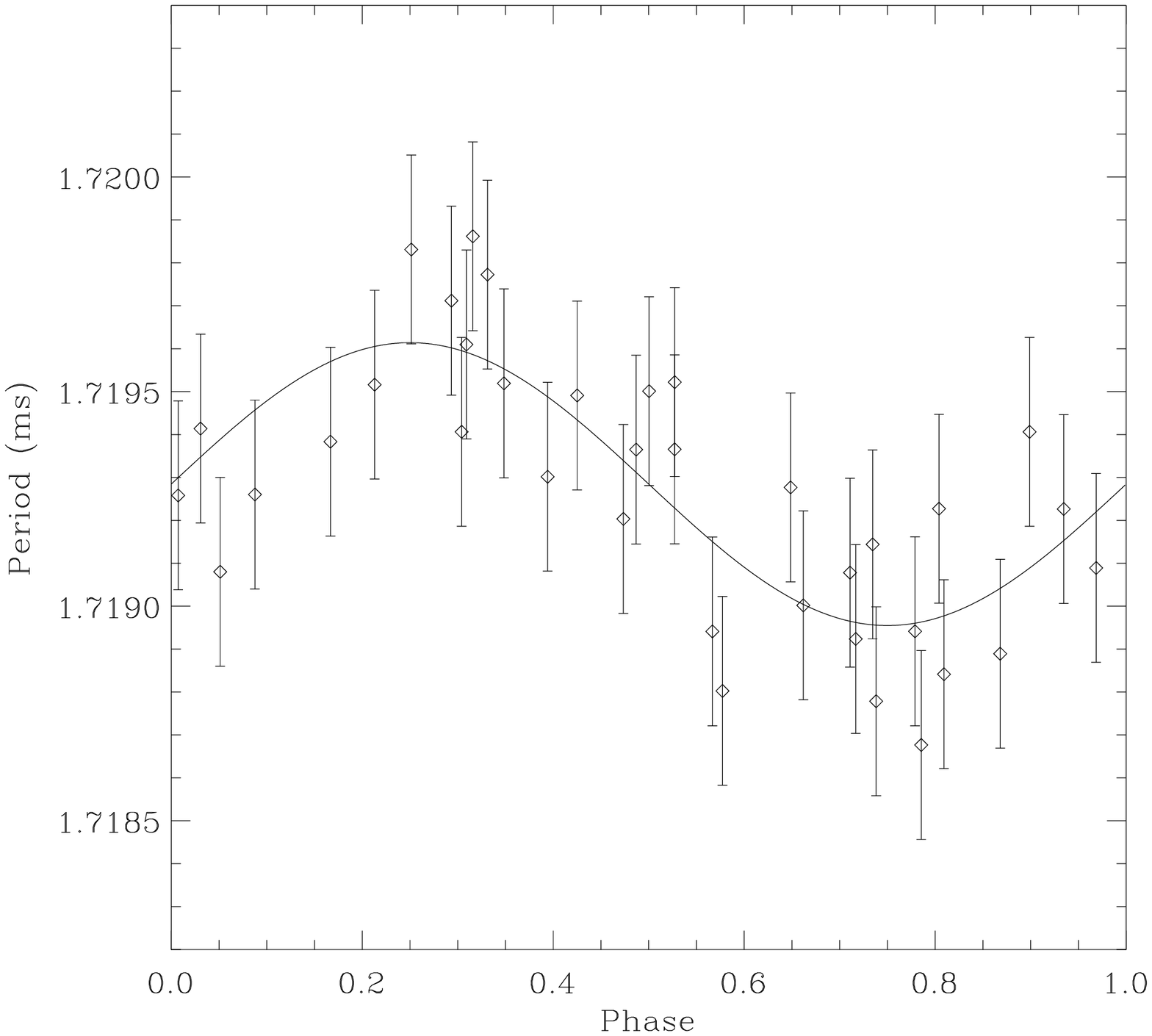]{Period versus orbital phase simulation 
using 36 simulated asymptotic periods sampled with the same statistical 
uncertainty as we estimated for our real measurements. We also show the best
fitting orbital doppler model. The neutron star velocity is detected at better
than $3\sigma$ confidence in this simulation. This suggests that a doubling of
the number of observed asymptotic periods should enable a detection of the 
neutron star velocity. \label{fig9}}

\newpage

\begin{figure}
\begin{center}
 \includegraphics[width=6in, height=7in]{f1.ps}
\end{center}

Figure 1: The {\it V\/} band light curves for 4U 1636-53 on 1998 
March 25 \& 27 and 1998 April 3. The solid sine curve marks our new ephemeris 
with a period of 3.70312064 hours. The dotted curve shows the ephemeris 
prediction of Augusteijn et al. (1998) for the same three nights with an 
arbitrary offset and amplitude. The number in each panel refers to the HJD 
starting at zero hours within each light curve.
\end{figure}
\clearpage

\begin{figure}
\begin{center}
 \includegraphics[width=6in, height=7in]{f2.ps}
\end{center}
Figure 2: The {\it V\/} band light curves for 4U 1636-53 on 1999 
June 9 and 2001 May 7 \& 8. The solid sine curve marks our new ephemeris 
and the first optical maximum on 7 May occurs at HJD = 2452036.954706 
The dotted curve shows the ephemeris prediction of Augusteijn et al. (1998) 
for the same three nights with an arbitrary offset and amplitude. The number 
in each panel refers to the HJD starting at zero hours within each light 
curve.
\end{figure}
\clearpage

\begin{figure}
\begin{center}
 \includegraphics[width=4in, height=3in]{f3a.ps}
 \hspace{2in}%
\includegraphics[width=4in, height=3in]{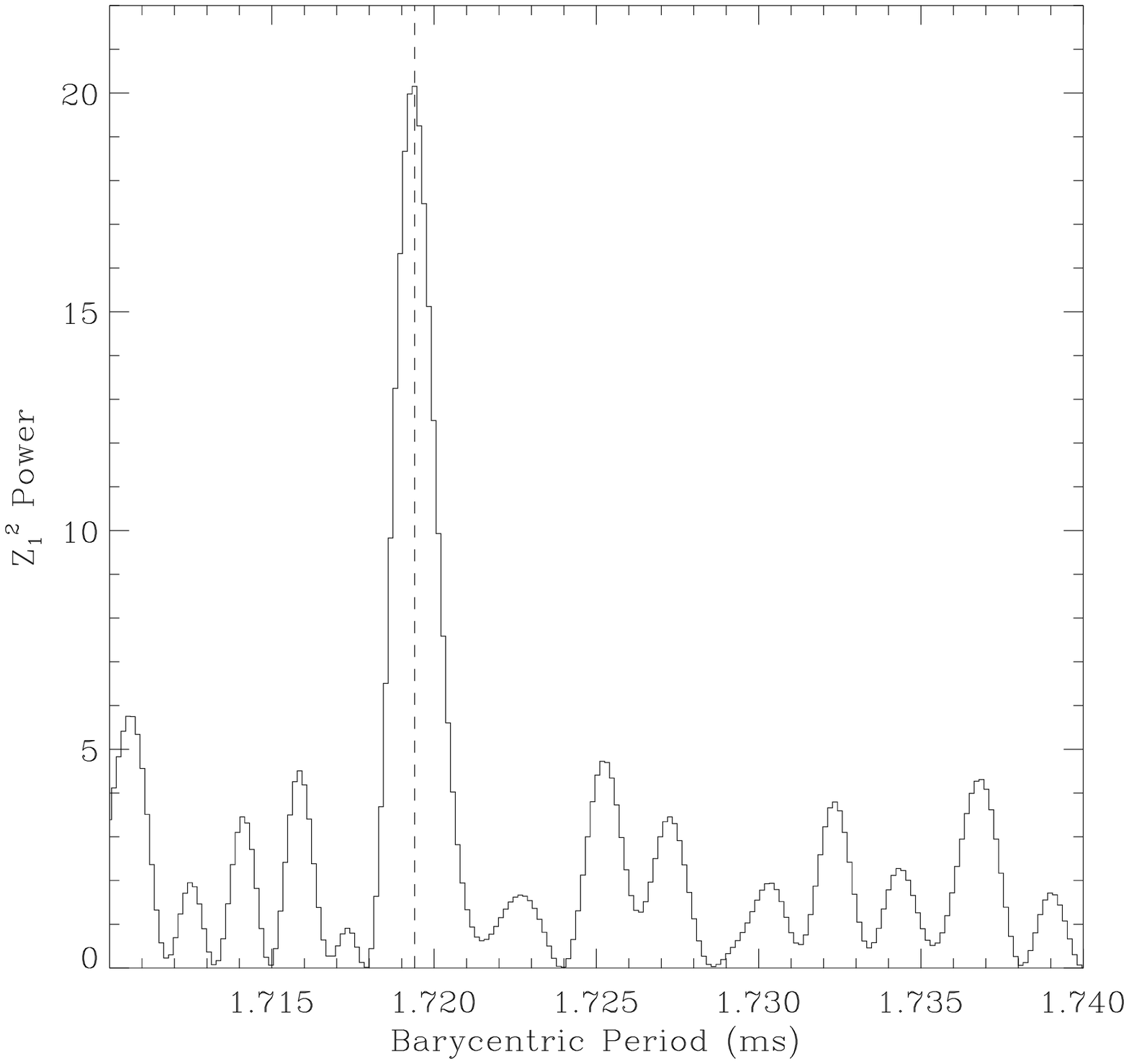}
\end{center}

Figure 3: Dynamic $Z_1^2$ spectrum in the hard X-ray
band (7 - 20 keV) for burst number 20 in Table 2 (top). The horizontal dashed 
line marks the asymptotic period inferred for this burst. The burst
lightcurve is overlaid (right axis). The gaps in the lightcurve are due to 
telemetry limitations for this data mode. Also shown is the $Z_1^2$ spectrum 
in the tail of the burst from which the asymptotic period was measured 
(bottom). In this case the vertical dashed line marks the asymptotic period. 
\end{figure}
\clearpage

\begin{figure}
\begin{center}
 \includegraphics[width=6in, height=6in]{f4.ps}
\end{center}
Figure 4: Histogram of simulated period measurements and
best fitting gaussian distribution. See the text (\S 4.1) for a discussion of 
the simulations. The width of the gaussian is $2.2\times 10^{-4}$ ms 
and represents the characteristic uncertainty in our asymptotic period 
measurements.
\end{figure}
\clearpage

\begin{figure}
\begin{center}
 \includegraphics[width=6in, height=6in]{f5.ps}
\end{center}
Figure 5: Histogram of measured asymptotic burst 
oscillation periods for 4U 1636-53. The periods are corrected to the solar 
system barycenter. Note the cluster of 18 periods centered near 1.7192 ms. 
A gaussian distribution centered at 1.71929 ms, of width 
$\sigma = 2.3\times 10^{-4}$ ms 
fits these data well and is shown by the thick solid curve. Note the presence 
of outliers towards longer period, but none shortward of the gaussian.
\end{figure}
\clearpage

\begin{figure}
\begin{center}
 \includegraphics[width=6in, height=6in]{f6.ps}
\end{center}
Figure 6: Plot of asymptotic period versus orbital
phase for the subset of 18 bursts which have a tightly clustered period
distribution.  The solid curve is the best fitting doppler model with
$v_{ns} \sin i = 18$ km s$^{-1}$. This fit, however, is not statistically
significant compared to a $v_{ns} \sin i = 0$ model (see \S 4.3). The dashed
curve shows the model with $v_{ns} \sin i = 55$ km s$^{-1}$, which is equal to
our $90\%$ confidence upper limit. The dotted curve shows the best fitting 
model when the phase of maximum is added as an additional parameter. This fit
has $v_{ns} \sin i = 59.3$ km s$^{-1}$.
\end{figure}
\clearpage

\begin{figure}
\begin{center}
 \includegraphics[width=6.5in, 
height=6in]{f7.ps}
\end{center}
Figure 7: Diagram of the Roche 
geometry for 4U 1636-53. The figure was drawn assuming neutron star and 
secondary masses of $1.6 M_{\odot}$ and $0.36 M_{\odot}$, respectively. 
The numbers circling the components correspond to orbital phases under the
assumption that phase zero corresponds to superior conjunction of the 
secondary (V801 Arae). The center of mass (CM) is denoted by a 
square symbol. 
The dashed circle around the neutron star marks the likely extent of an 
accretion disk under the assumption that it fills $90\%$ of the neutron star
Roche lobe. The triangles and shaded region mark the inferred locations of
the radial velocity components measured by Augusteijn et al. (1998).
\end{figure}
\clearpage

\begin{figure}
\begin{center}
 \includegraphics[width=6.0in, 
height=6in]{f8.ps}
\end{center}
\vskip 20pt
Figure 8: Constraints on the component masses in 
4U 1636-53 derived from our upper limit on $v_{ns} \sin i$ and the radial
velocity data of Augusteijn et al. (1998). The regions with horizontal hatching
are excluded by the neutron star velocity limit, while the vertical hatched
regions are excluded by the radial velocity data. We show constraints for
$i = 50^{\circ}$ and $v_{ns} \sin i < 55$ km s$^{-1}$ (a), $i = 50^{\circ}$ 
and $v_{ns} \sin i < 75$ km s$^{-1}$ (b), $i = 60^{\circ}$ and $v_{ns} \sin i 
< 55$ km s$^{-1}$ (c), $i = 60^{\circ}$ and $v_{ns} \sin i < 75$ km s$^{-1}$ 
(d). The thick lines denote $M_{ns} = 1.4 M_{\odot}$ and 
$M_{sec} = 0.36 M_{\odot}$, respectively. See \S 5.1 for a discussion on how
the constraints were derived.
\end{figure}
\clearpage

\begin{figure}
\begin{center}
 \includegraphics[width=6.0in, 
height=6in]{f9.ps}
\end{center}
Figure 9: Period versus orbital phase simulation 
using 36 simulated asymptotic periods sampled with the same statistical 
uncertainty as we estimated for our real measurements. We also show the best
fitting orbital doppler model. The simulation used $v\sin i = 55$ km s$^{-1}$,
our $90\%$ confidence upper limit. The neutron star velocity is detected at 
better than $3\sigma$ confidence in this simulation. 
This roughly suggests that a doubling of the number of observed asymptotic 
periods should enable a detection of the neutron star velocity. We note, 
however, that due to the fact that only $\sim 70\%$ of bursts fall within the 
asymptotic distribution, this would correspond to more than a doubling of the
current burst sample. 
\end{figure}
\clearpage

\begin{table*}
\begin{center}{Table 1: Optical observations of 4U 1636-53}
\begin{tabular}{ccccccc} \\
\tableline
\tableline
Date       &  HJD Start    &  HJD End      &  Filter  &  Int.  & No.   \\
           &  -2450000     &  -2450000     &          &  (s)   & Exp.  \\
\tableline
  3/25/98  &  0898.01619  &  0898.30426  &  {\it V\/} \& {\it I\/}  &  180  &  
65  \\
  3/27/98  &  0899.99518  &  0900.23595  &  {\it V\/} \& {\it I\/}  &  180  &  
56  \\
  4/ 3/98  &  0907.09065  &  0907.32071  &  {\it V\/}  &  300  &  65  \\
  3/26/99  &  1264.02547  &  1264.30913  &  {\it V\/}  &  300  &  32  \\
  3/28/99  &  1266.15552  &  1266.26166  &  {\it V\/}  &  300  &  10  \\
  3/31/99  &  1269.15566  &  1269.29098  &  {\it V\/}  &  300  &  16  \\
  4/ 2/99  &  1271.08660  &  1271.32108  &  {\it V\/}  &  300  &  33  \\
  4/ 4/99  &  1273.25977  &  1273.31819  &  {\it V\/}  &  300  &   8  \\
  6/ 9/99  &  1338.88939  &  1339.14694  &  {\it V\/}  &  180  &  81  \\
  6/10/99  &  1340.18839  &  1340.27722  &  {\it V\/}  &  180  &  12  \\
  5/ 7/01  &  2036.89853  &  2037.34667  &  {\it V\/}  &  300  &  72  \\
  5/ 8/01  &  2037.97337  &  2038.28175  &  {\it V\/}  &  300  &  74  \\
\tableline
\end{tabular}
\end{center}
\end{table*}
\clearpage
\begin{table*}
\begin{center}{Table 2: X-ray bursts detected from 4U 1636-53 by RXTE}
\begin{tabular}{ccccccc} \\
\tableline
\tableline
Burst  & RXTE     &  Date & HJD      &  Period         &  Binary      \\
No.    & Obs. ID. &       & -2450000 &  (ms)           &  phase       \\
       &          &       &          &  $\pm$ 0.00040  &  $\pm$ 0.05  \\
\tableline
 1  &  10088-01-07-02   &  12/28/96  &  0446.439466  &  1.71940  &  0.42  \\
 2  &  10088-01-07-02   &  12/28/96  &  0446.491308  &  1.71910  &  0.75  \\
 3  &  10088-01-08-01   &  12/29/96  &  0447.472404  &  1.71925  &  0.96  \\
 4  &  10088-01-08-030  &  12/31/96  &  0449.229474  &        -  &  0.07  \\
 5  &  10088-01-09-01   &   2/23/97  &  0502.913912  &  1.72028  &  0.75  \\
 6  &  30053-02-02-02   &   8/19/98  &  1044.991053  &  1.72083  &  0.60  \\
 7  &  30053-02-01-02   &   8/20/98  &  1045.654542  &  1.72161  &  0.80  \\
 8  &  30053-02-02-00   &   8/20/98  &  1045.719849  &  1.71954  &  0.22  \\
 9  &  40028-01-02-00   &   2/27/99  &  1236.865609  &  1.71954  &  0.64  \\
10  &  40028-01-04-00   &   4/29/99  &  1297.575867  &  1.71925  &  0.77  \\
11  &  40028-01-06-00   &   6/10/99  &  1339.751875  &  1.72240  &  0.63  \\
12  &  40028-01-08-00   &   6/18/99  &  1348.493173  &  1.71943  &  0.94  \\
13  &  40030-03-04-00   &   6/19/99  &  1349.234723  &  1.72260  &  0.63  \\
14  &  40031-01-01-06   &   6/21/99  &  1351.300601  &  1.71930  &  0.70  \\
15  &  40028-01-10-00   &   9/25/99  &  1447.360320  &        -  &  0.49  \\
16  &  40028-01-13-00   &   1/22/00  &  1565.570419  &        -  &  0.44  \\
17  &  40028-01-13-00   &   1/22/00  &  1565.703136  &  1.72043  &  0.28  \\
18  &  40028-01-14-01   &   1/30/00  &  1573.506255  &        -  &  0.65  \\
19  &  40028-01-15-00   &   6/15/00  &  1710.717286  &  1.72147  &  0.82  \\
20  &  40028-01-18-000  &   8/ 9/00  &  1765.557103  &  1.71940  &  0.80  \\
21  &  40028-01-18-00   &   8/ 9/00  &  1765.875286  &  1.71880  &  0.81  \\
22  &  40028-01-19-00   &   8/12/00  &  1769.482989  &  1.71940  &  0.64  \\
\tableline
\end{tabular}
\end{center}
\end{table*}

\pagebreak
\begin{table*}
\begin{center}{Table 2 (cont.): X-ray bursts detected from 4U 1636-53 by RXTE}
\begin{tabular}{ccccccc} \\
\tableline
\tableline
Burst  & RXTE     &  Date & HJD      &  Period         &  Binary      \\
No.    & Obs. ID. &       & -2450000 &  (ms)           &  phase       \\
       &          &       &          &  $\pm$ 0.00040  &  $\pm$ 0.05  \\
\tableline
23  &  40028-01-20-00   &  10/03/00  &  1821.479052  &  1.71969  &  0.63  \\
24  &  50030-02-01-00   &  11/05/00  &  1853.677829  &  1.72043  &  0.36  \\
25  &  50030-02-02-00   &  11/12/00  &  1861.247296  &  1.71971  &  0.26  \\
26  &  50030-02-04-00   &   1/28/01  &  1937.613090  &  1.71968  &  0.44  \\
27  &  50030-02-05-01   &   2/01/01  &  1942.372888  &  1.71980  &  0.56  \\
28  &  50030-02-05-00   &   2/02/01  &  1942.597558  &  1.71932  &  0.98  \\
29  &  50030-02-09-000  &   4/05/01  &  2005.215668  &  1.71910  &  0.18  \\
30  &  50030-02-10-00   &   4/30/01  &  2029.732151  &  1.71907  &  0.30  \\
\tableline
\end{tabular}
\end{center}
\end{table*}

\end{document}